\documentclass[aps,prl,amsmath,amssymb,amsbsy,twocolumn,superscriptaddress]{revtex4-1}

\usepackage{amsfonts} 
\usepackage{amsmath}
\usepackage{amssymb}
\usepackage{cancel} 
\usepackage{bm}
\usepackage[usenames, dvipsnames]{color} 
\usepackage{dcolumn}
\usepackage{epic} 
\usepackage{epsfig}
\usepackage{eucal} 
\usepackage{feynmp}    
\usepackage{graphicx}
\usepackage[breaklinks,colorlinks = true,linkcolor = red,urlcolor=blue,citecolor=red]{hyperref}
\usepackage{mathrsfs}
\usepackage{soul}
\usepackage{subfigure}
\usepackage{wrapfig} 
\usepackage{xy} 

\begin{document}

\title{Light-cone  spreading of perturbations and the butterfly effect in a classical spin chain}


\author{Avijit Das}
\email{avijit.das@icts.res.in}
\affiliation{International Centre for Theoretical Sciences, Tata Institute of Fundamental Research, Bengaluru 560089, India}
\author{Saurish Chakrabarty}
\thanks{Present address : School of Chemical and Biomedical Engineering, Nanyang Technological University, Singapore 637459.}
\affiliation{International Centre for Theoretical Sciences, Tata Institute of Fundamental Research,  Bengaluru 560089, India}
\author{Abhishek Dhar}
\affiliation{International Centre for Theoretical Sciences, Tata Institute of Fundamental Research,  Bengaluru 560089, India}
\author{Anupam Kundu}
\affiliation{International Centre for Theoretical Sciences, Tata Institute of Fundamental Research,  Bengaluru 560089, India}
\author{David A. Huse}
\affiliation{Physics Department, Princeton University, Princeton, NJ 08544, USA}
\author{Roderich Moessner}
\affiliation{Max-Planck Institute for the Physics of Complex Systems, 01187 Dresden, Germany}
\author{Samriddhi Sankar Ray}
\affiliation{International Centre for Theoretical Sciences, Tata Institute of Fundamental Research,  Bengaluru 560089, India}
\author{Subhro Bhattacharjee}
\affiliation{International Centre for Theoretical Sciences, Tata Institute of Fundamental Research,  Bengaluru 560089, India}

\begin{abstract}
We find that the effects of a localised perturbation in a  chaotic classical many-body system--the classical  Heisenberg  chain at infinite temperature--spread ballistically with a finite speed even when the local spin dynamics is diffusive. We study two complementary aspects of this butterfly effect: the rapid growth of the perturbation, and its simultaneous ballistic (light-cone) spread, as characterised by the Lyapunov exponents and the butterfly speed respectively. We connect this to recent studies of  the out-of-time-ordered commutators (OTOC), which have been proposed as an indicator of chaos in a quantum system. We provide a straightforward identification of the OTOC with a natural correlator in our system and demonstrate that many of its interesting qualitative features  are present in the classical system.  Finally, by analysing the scaling forms, we relate the growth, spread and propagation of the perturbation with the growth of one-dimensional interfaces described by the Kardar-Parisi-Zhang (KPZ) equation. 
\end{abstract}
\date{\today}
\maketitle
\paragraph*{Introduction :}  The {butterfly effect} \cite{lorenz1996essence, hilborn2004sea, lorenz2000butterfly} is a vivid picture for the sensitivity of a spatially extended chaotic many-body system  to arbitrarily small changes to its initial conditions. In this picture, this exquisite sensitivity -- the proverbial butterfly wingbeat is enough to make the difference between presence or absence of a tornado -- perhaps takes  precedence over the fact that these changes are global --  tornado activity is  toggled in a  place far away from the butterfly.  While this sensitivity to initial conditions  is well-studied and quantified via the (positive) Lyapunov exponents, the spatial spreading of the perturbation has received somewhat less attention.  This spreading, if ballistic, is characterised by a {\it butterfly speed}.  Lyapunov exponents and butterfly speed thus encode two complementary aspects of the butterfly effect. 

These issues have acquired additional interest in the context of many recent studies of scrambling of information in quantum many body systems \cite{aleiner2016microscopic, sekino2008fast, Shenker:2013pqa,brown2012scrambling,Lashkari2013, maldacena2016bound, rozenbaum2017lyapunov, tsuji2017bound, iyoda2017scrambling,kukuljan2017weak, kurchan2016quantum, roberts2016lieb, swingle2016measuring, swingle2017slow, bohrdt2016scrambling, stanford2016many, banerjee2017solvable, Shenker2014, Hosur2016}. In this setting, the out-of-time-ordered  commutator (OTOC) \cite{1969LarkinOvchin, kitaev}  has emerged as a diagnostic  \cite{sekino2008fast,1969LarkinOvchin, maldacena2016bound, Shenker:2013pqa,brown2012scrambling,Lashkari2013, maldacena2016bound, rozenbaum2017lyapunov, tsuji2017bound, iyoda2017scrambling,kukuljan2017weak, kurchan2016quantum, roberts2016lieb, swingle2016measuring, swingle2017slow, bohrdt2016scrambling, stanford2016many, banerjee2017solvable, Shenker2014, Hosur2016}: for two Hermitian operators $\hat{W}(x,t)$ and $\hat{V}(0,0)$ localised around $x$ at time $t$ and $x=0$ at time $t=0$ respectively, the OTOC,  defined as $F(t)=-\langle [\hat{W}(x,t),\hat{V}(0,0)]^2 \rangle $, estimates the effect of the operator, $V(0,0)$ on the measurement of operator, $W(x,t)$.  In a class of large $N$ gauge theories it was found that, for a given $x$ and $t$,  the OTOC is generically characterised by an exponent $\tilde\lambda$, and a velocity $\tilde v_{\rm B}$, which are respectively the measure of the exponential growth and the speed of spreading of the initially localised perturbation. Analogous to classical dynamical systems, the former is often identified with the largest Lyapunov exponent, and the latter with the butterfly speed. 

Interestingly, these twin features are present even when the usual probes for relaxation and equilibration in a many-body system, the two-point functions $\langle \hat{W}(t) \hat{V}(0) \rangle$, are diffusive and hence do not capture the above ballistic  spread. This was observed in a study of the OTOC in a system with diffusive energy transport-- the one-dimensional Bose-Hubbard chain \cite{bohrdt2016scrambling, PhysRevB.96.054503}  and diffusive metals \cite{PhysRevX.7.031047} at finite temperature and also in the context of random unitary circuits \cite{khemani2017operator, rakovszky2017diffusive}, which lend themselves to a considerable degree of analytical and numerical insight \cite{von2017operator,nahum2017quantum,nahum2017operator}.

In this paper, we present a detailed analysis of the spatio-temporal evolution of the divergence of the dynamical trajectories of perturbed and unperturbed systems. Our model is a well-known classical many-body system--the Heisenberg spin-chain at high temperatures, 
whose classical Hamiltonian dynamics of the spins is diffusive. We first identify a correlator which represents a natural classical limit of an OTOC, and turns out to be a very simple quantity: the decrease of the correlation between the system and its perturbed copy under time evolution. In particular, we find that the divergence of dynamical trajectories spreads in space ballistically. We provide an accurate extraction of the corresponding Lyapunov exponent and butterfly speed, and provide a description of the variations in the divergence between different initial states in terms of a KPZ-based analysis, which yields scaling forms for the distributions.

Our work connects to earlier studies of the propagation of chaos on coupled map lattices with discrete time evolution \cite{kanekobook, lepri1996chronotopic}, partial differential equations \cite{vastano1988information,gaspard1998experimental,grassberger1999p} and anharmonic coupled oscillator chains  \cite{giacomelli2000convective}, where the concept of a velocity-dependent Lyapunov exponent was formulated \cite{deissler1987velocity,kanekobook,lepri1997chronotopic}  and  related to the speed of spread of correlations \cite{giacomelli2000convective}.  In parallel, the concrete classical limit of the OTOC provides a natural platform to investigate the existence and nature of intrinsic differences in  spatio-temporal chaos between classical and quantum many-body systems \cite{scaffidi2017semiclassical, khemani2018velocity, nussinov2014thermalization}.

\begin{figure}
\centering
\includegraphics[scale=0.23]{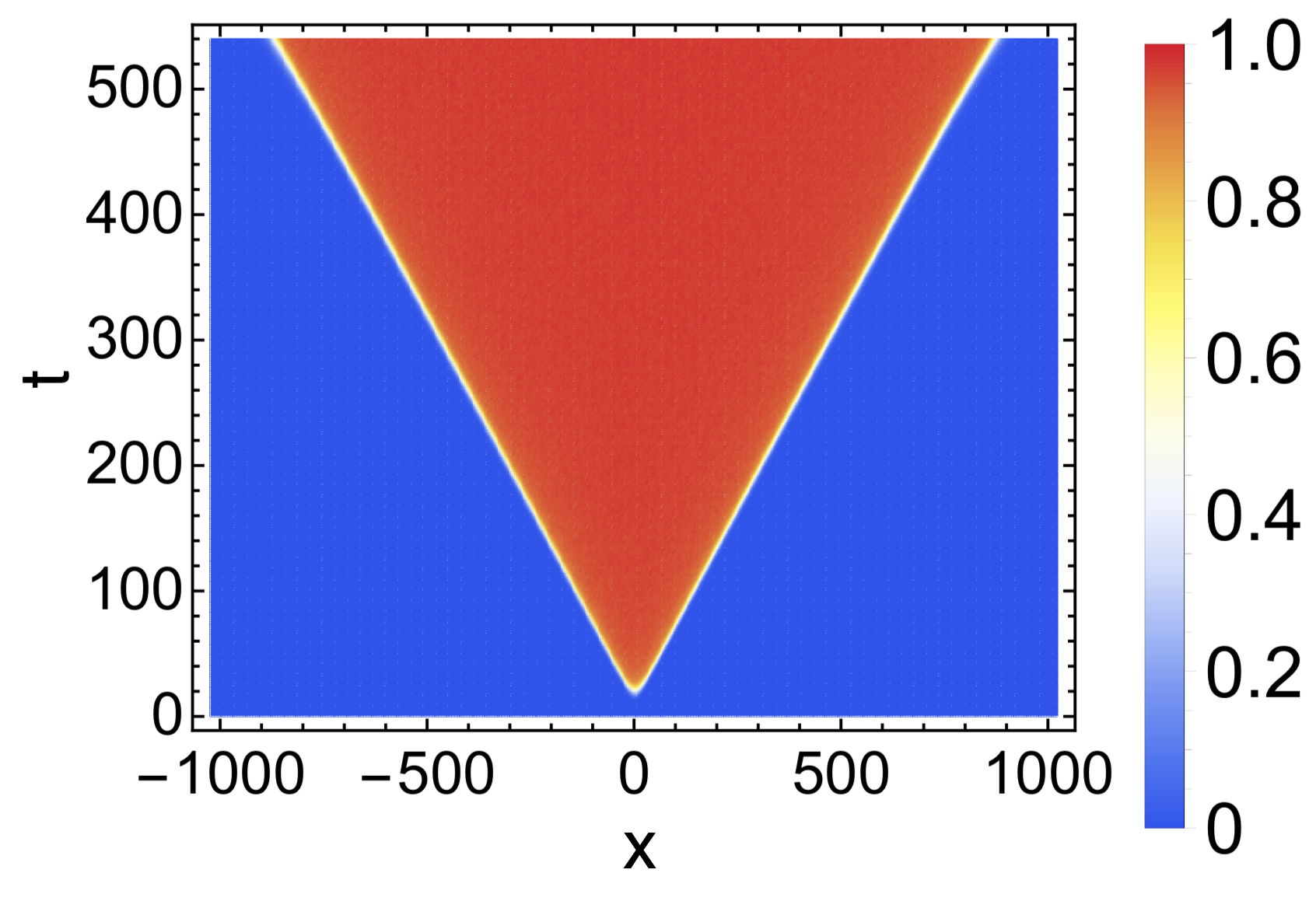}
\caption{(Color online) Simultaneous growth and ballistic spread of a perturbation in a classical Heisenberg spin chain whose spin dynamics (Fig. \eqref{figdiff}) is diffusive at $T=\infty$. The speed of spreading obtained from the classical OTOC, $D(x,t)$ (see text), defines a ``light cone''.  The results are shown for a  perturbation at time $t=0$ of size  $\varepsilon=10^{-3}$ at the centre of a system of size $L=2048$.}
\label{figlightcone}
\end{figure}

\paragraph*{The Heisenberg spin chain:} We consider a one-dimensional lattice of spins ${\bf S}_x~, x=0,\ldots,N-1$ described by the Heisenberg  Hamiltonian 

\begin{align}
\mathcal{H}=-J\sum_{x=0}^{N-1} {\bf S}_x\cdot{\bf S}_{x+1}~,
\end{align}
 where $J>0$ and ${\bf S}_x$ are unit three component classical vectors and we take periodic boundary conditions ${\bf S}_x\equiv{\bf S}_{x+N}$. We consider  a classical precessional dynamics  
\begin{equation}
\frac{d{\bf S}_x}{dt}=J{\bf S}_x\times ({\bf S}_{x-1}+{\bf S}_{x+1})=\{ {\bf S}_x,H\}~, \label{eom}
\end{equation}
where the spin-Poisson bracket is defined as $\{f,g\}= \sum_x \sum_{\alpha,\beta,\gamma} \epsilon^{\alpha\beta\gamma}S_x^\gamma (\partial f/\partial {S}_x^\alpha) (\partial g/\partial {S}_x^\beta) $ for arbitrary functions $f,g$ of the spin variables.

\begin{figure}
 \centering
\includegraphics[scale=0.25]{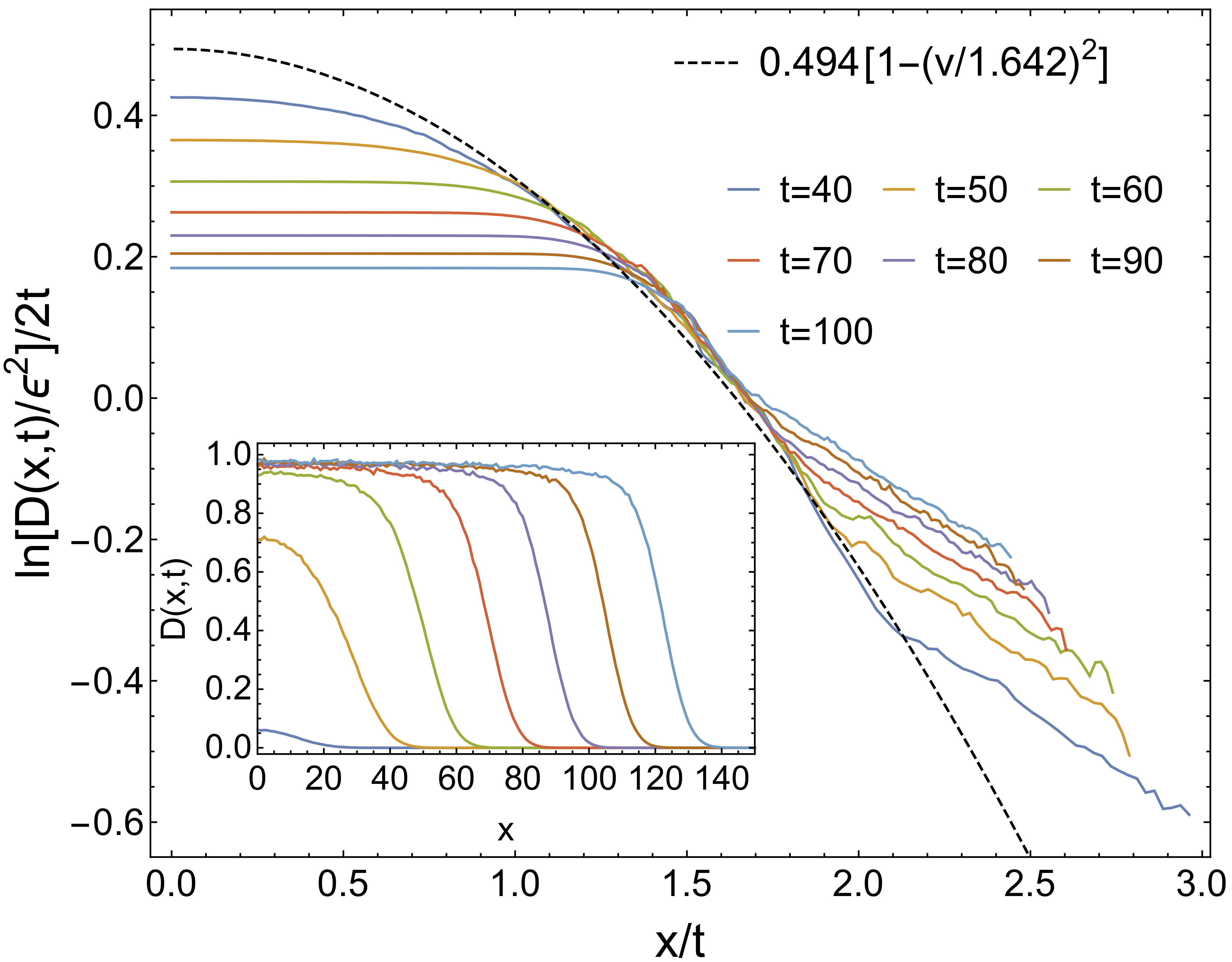}
\caption{(color onine) The inset plots  $D(x,t)$ as a function of $x$, at different times ($t=40,50,\cdots 100$), showing growth and ballistic  propagation of the perturbtion front. The scaled data (main panel) shows that the front is fit well by Eq.~(\eqref{Dform}) with $\mu=0.494$ and $v_b=1.642$ near $x\sim v_b t$.  Here $\epsilon=10^{-8}$ and averaging was done over $2\times 10^4$ realizations.}
\label{DxtVsx}
\end{figure}

\paragraph*{Classical OTOC analogue :}
We consider  two spin configurations which, at $t=0$, differ {\it only} at site $x=0$ by a  rotation, $\varepsilon$, that is either small or infinitesimal, about an axis $\hat{\bf n}=(\hat{\bf z}\times {\bf S}_0)/|\hat{\bf z}\times {\bf S}_0|$ (where $\hat{\bf z}$ is the unit vector along the global $z$-axis) such that $\delta {\bf S}_0=\varepsilon (\hat{\bf n} \times {\bf S}_0)$.  We  study the spreading of such a localised perturbation. For infinitesimal $\varepsilon$, the change at some point $x$ is given by $\delta S_x^\alpha(t) \approx (\partial S_x^\alpha(t)/\partial S_0^\beta) \delta S_0^\beta= \varepsilon~n^\gamma\epsilon^{\beta \gamma \nu} S_0^\nu ~(\partial S_x^\alpha(t)/\partial S_0^\beta) =\varepsilon~n^\gamma \{S_x^\alpha(t),S_0^\gamma(0) \}$. To measure the evolution of the perturbation we define
\begin{equation}
2 D(x,t) := \langle (\delta {\bf S}_x(t))^2 \rangle \approx \varepsilon^2 \langle \{ {\bf S}_x(t),\hat{\bf n}\cdot{\bf S}_0 \}^2 \rangle. \label{Ddefn}
\end{equation}
where, throughout this paper, $\langle\cdots\rangle$ denotes averaging over spin configurations chosen from the equilibrium distribuition $P(\{ {\bf S}_x \})=e^{-\mathcal{H}/T}/Z(T)$ and $Z(T)$ is the partition function. Denoting the two initial spin configurations discussed above by $\{ {\bf S}^a_x(t=0)\}$ and $\{ {\bf S}^b_x(t=0)\}$, we can obtain a simpler expression as
\begin{equation}
D(x,t)=1-\langle {\bf S}^a_x(t)\cdot{\bf S}^b_x(t)\rangle. \label{Dreln}
\end{equation}
where $\langle {\bf S}^a_x(t)\cdot{\bf S}^b_x(t)\rangle$ is the {\it cross-correlator} between the two copies. If the dynamics is chaotic, as is known to be in this classical spin-chain at infinite temperatures \cite{de2012largest,das2002driven}, we expect that for any $x \neq 0$, the above  quantity, as a function of time, $t$, starts from the value $0$ (when the spins of the two copies at a given $x$ are perfectly correlated) and asymptotes to $1$ (when they are completely de-correlated).   {\it Thus $D(x,t)$ indeed measures the spatio-temporal evolution of de-correlation throughout the system.} Apart from $D(x,t)$, we also calculate the usual dynamic spin-correlation function
\begin{equation}
C(x,t)=\langle {\bf S}_x(t)\cdot {\bf S}_0(0) \rangle~. 
\label{eq_cxt}
\end{equation}

At this point, it is useful to understand the connection between $D(x,t)$ and the OTOC. On canonical quantisation of the theory obtained by replacing the Poisson bracket with the commutator, {\it i.e.} $\{f,g\}\rightarrow\frac{1}{i\hbar}[f,g]$, we get ${D}(x,t)\rightarrow -\frac{\varepsilon^2}{\hbar^2}Tr\left[\rho_T([{\bf S}_x(t),\hat{\bf n}\cdot{\bf S}_0(0)])^2\right]$,  where ${\bf S}_x$ are now quantum operators. This is nothing but the finite temperature generalisation of the OTOC introduced earlier with $\hat W(x,t)={\bf S}_x(t)$  and $\hat V(0,0)=\varepsilon\hat{\bf n}\cdot {\bf S}_0(0)$.


\paragraph*{Numerical Results:} We now present representative results of our numerical simulation   of the Heisenberg spin chain with periodic boundary conditions. The simulations were performed using a fourth-order Runge-Kutta (RK4) numerical integration scheme for the spin dynamics. For the numerical simulations,  energy is measured in units of $J$. The time-step in RK4 was taken to be $\Delta t=0.001-0.005$  such that the energy/site and magnetisation/site were conserved up to $\sim 10^{-12}$.  The configuration averaging was done over $\sim 10^{5}$ equilibrated initial conditions for $C(x,t)$ and $\sim 10^{4}$ for $D(x,t)$. Many of the simulations had to be performed at quadruple level machine precision.

\begin{figure}
\includegraphics[scale=0.29]{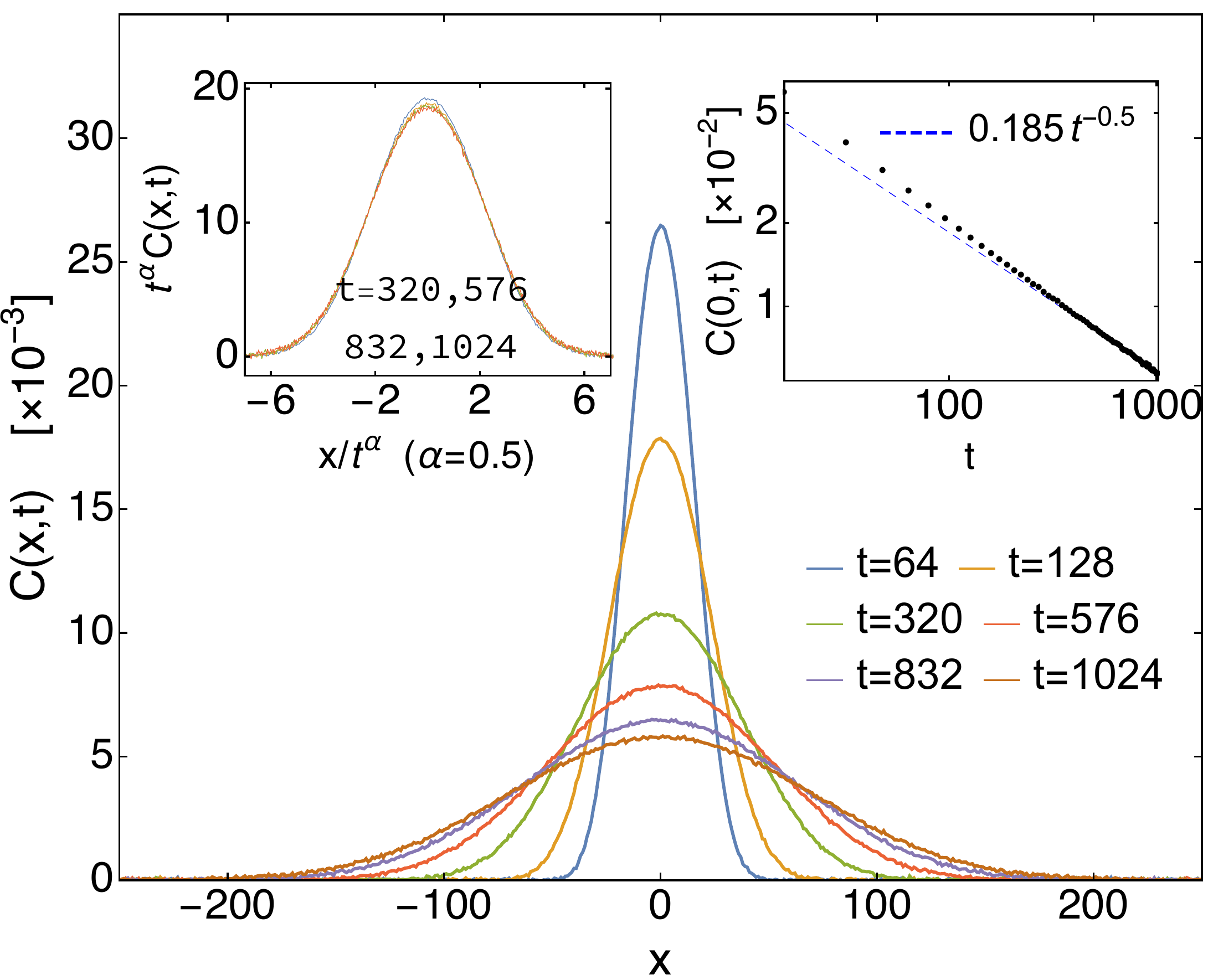}
\caption{(Color online) The spatial profile of $C(x,t)$ (Eq. \eqref{eq_cxt}) at different times, $t$, for a system of size $L=512$ at $T=\infty$ with averaging over $10^{5}$ initial conditions. The left inset shows a collapse of the data after a diffusive scaling of $x/\sqrt{t}$ while the right inset shows the resultant $t^{-1/2}$ scaling of the auto-correlation.}
\label{figdiff}
\end{figure} 

Our first main finding, namely ballistic propagation of the de-correlation,  is summarised in Fig.~\eqref{figlightcone} which shows that the OTOC falls sharply outside  a light cone. The light cone is specified by the lines $x=\pm v_b t$ where $v_b$ is the butterfly speed. For the two systems whose de-correlation $D(x,t)$ measures,  the red region in Fig.~\eqref{figlightcone} corresponds to complete de-correlation with $\langle{\bf S}^a_x(t)\cdot{\bf S}^b_x(t)\rangle \cong 0$. This also gives the natural definition of the {\it light-cone} velocity in the sense of a ``classical Lieb-Robinson speed" \cite{lieb1972finite,PhysRevLett.112.210601, Marchioro1978} which is then equal to the butterfly speed. 

\begin{figure}
\centering
\includegraphics[scale=0.47]{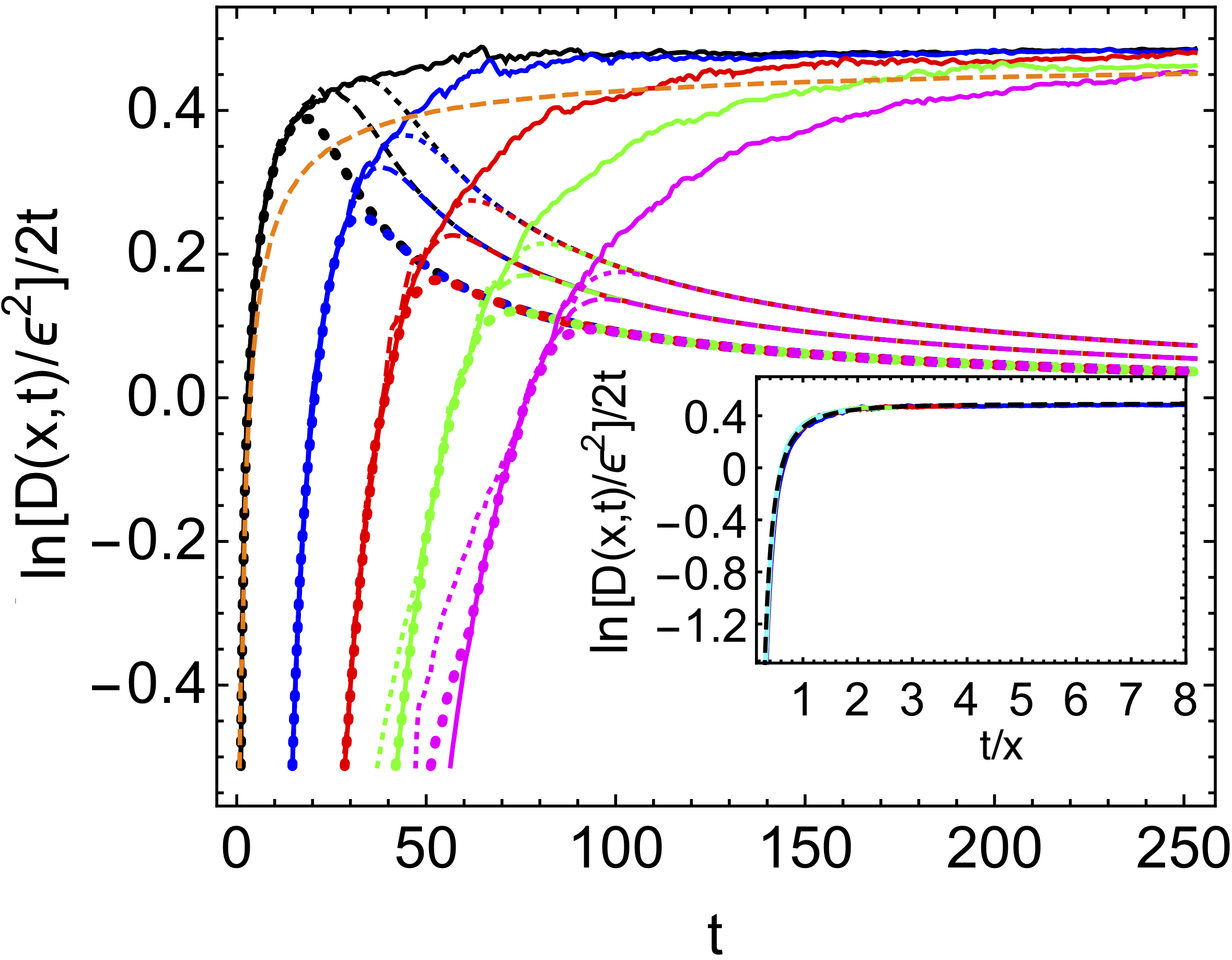}
\caption{
(Color online) (a) Plot of $\ln [D(x,t)/\varepsilon^2]/(2t)$ versus $t$ at $x=0$ (black), $32,64,96$,$128$(magenta), for $\varepsilon=10^{-4}$ (thick dotted lines), $\varepsilon=10^{-6}$(dashed lines) and $\varepsilon=10^{-8}$ (thin dotted lines), for $L=1024$  from  solving the non-linear Equation of motion (Eq. \eqref{eom}). The solid lines are results from the linearized dynamics and correspond to the limit $\epsilon \to 0$ and hence gives $\lambda_D$ (see text). The dashed orange line corresponds to   $\langle \ln [\delta {\bf S}^2(x,t)/2\varepsilon^2]\rangle /(2t)$, obtained  from the linearized dynamics for $x=0$ and we see the slightly different saturation value corresponding to $\lambda_L$ (see text). (b) Inset plots the results for the linearized dynamics for $x=32,64,96,128$ on scaling the time axis by $x$. The collapsed data approximately fits the solid line corresponding to the scaling form Eq.~\eqref{Dform} with $\mu\approx0.494, v_b\approx1.64$.}
\label{DxtVst}
\end{figure}

In Fig.~\eqref{DxtVsx} we  plot the signal $D(x,t)$ at different times to show the propagation of the front. As can be seen from the scaling, the front (for $x\sim v_b t$) is fit well by  
\begin{equation}
 D(x,t)=\varepsilon^2 \exp {[2 \mu t( 1- (x/v_b t)^2 )]}~, \label{Dform}
\end{equation}
with $\mu \approx 0.494,~v_b \approx 1.6417(2)$.  The deviations in scaling seen for $ x \sim v_b t$ arise from errors due to finite machine precision (quadruple level precision in this case). Later [see Fig. (\ref{DxtVst})] we shall see that working with a linearized dynamics avoids these errors and we get much better collapse of data in the entire range. The scaling function is quite accurate within the light cone but in general is  only an approximate fit for $x\gtrsim v_b t$.  The finite butterfly speed is in stark contrast with the entirely diffusive \cite{gerling1990time} spin dynamics as recorded by the regular two point correlator $C(x,t)$ (Eq. \eqref{eq_cxt}) shown in Fig. \eqref{figdiff}.   The characteristic signature of diffusion-- $x/\sqrt{t}$ collapse at long times-- is clearly visible in the insets of Fig.~\eqref{figdiff}. 
   
\begin{figure}
 \centering
\includegraphics[scale=0.5]{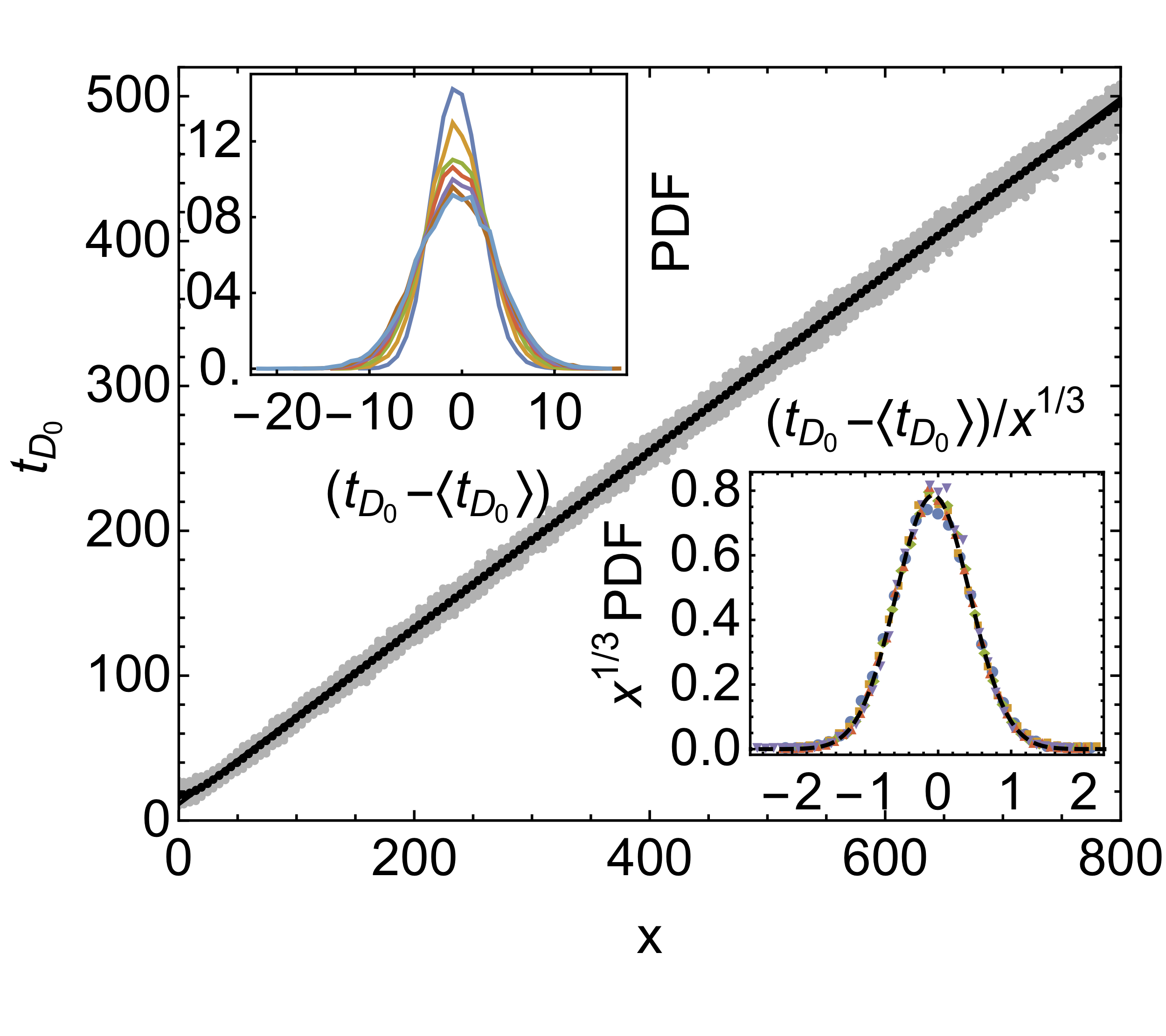}
\caption{(color online) The main panel shows $t_{D_0}$ (defined in the main text) as a function of $x$, for $D_0=100 \epsilon=0.1$ and different initial spin-configurations (grey scatter). The mean (black connected data-points) over $10^4$ configurations is also shown and has a slope $\approx 1/(1.6417(2))$.  The upper inset 
shows the distribution of $t_{D_0}$ at space-points $x=100,200,\ldots,700$ while the lower inset shows collapse of the distributions  with a width scaling as $\sim x^{1/3}$. The dotted curve in the lower inset is the gaussian fit to the fluctuations at $x=600$.}
\label{tDvsx}
\end{figure}

An alternate way of analysing the data is to ask at what time, $t_{D_0}$,  the signal attains a threshold value $D_0$ at a given $x$ for a set of different realisation of random initial configurations. In Fig.~\eqref{tDvsx} we plot the resulting set of $t_{D_0}$'s as a function of $x$. Its mean grows as $t_{D_0} =x/v_b$, with $v_b\approx 1.64$ in accordance with Fig. \eqref{DxtVsx}. Importantly, there is a spread of times for the arrival of the front leading to a distribution of times $t_{D_0}$ for a given $x$.  This distribution for different values of $x$ as well as its collapse indicating a $x^{2/3}$ scaling of variance of $t_{D_0}$ is shown in the inset of Fig. \eqref{tDvsx}.  Thus there are variations between different initial states in the timing of the front's arrival that are of order $\sim x^{1/3}$.

We next analyse the properties of the front in more detail, starting with its   exponential growth in the temporal regime and then considering  its fluctuations within a KPZ framework.  From the usual definition of the Lyapunov exponents, we expect  the quantity $\lim_{\epsilon \to 0} \delta S_x(t)^2/\epsilon^2$ to grow exponentially with time (at large, but finite times) as $\sim e^{2 \lambda({\bf S},t) t}$, for any $x$, where the Lyapunov exponent at time $t$, $\lambda({\bf S},t)$, may depend on the initial spin-configuration $\{{\bf S}\}$ of a given realisation. In the limit $\varepsilon\rightarrow 0$, it is possible to write the linearised equation of motion for $\lim_{\epsilon \to 0} \delta {\bf S}_x:={\bf z}_x$, 
\begin{equation}
\dot{{\bf z}}_x=J {\bf S}_x \times ({\bf z}_{x-1}+{\bf z}_{x+1}) + J {\bf z}_x \times ({\bf S}_{x-1}+{\bf S}_{x+1})~,
\label{lindyn} 
\end{equation}
where ${\bf S}$, obtained by solving the equation of motion Eq.~\eqref{eom} for a given random initial configuration,  acts as the dynamic field for ${\bf z}$. The linearised equation can then be used to obtain the Lyapunov exponent.   By sampling random initial configurations, we  can then define an average exponent  $\lambda_L(t)=\langle \lambda ({\bf S},t) \rangle$.  Given (from Eq. \eqref{Dreln}), $D(x,t)=\langle\left(\delta{\bf S}_x(t)\right)^2\rangle/2$, we expect  $\lim_{\epsilon \to 0} D(x,t)/\epsilon^2$   to grow exponentially with time as $\sim e^{2 \lambda_D(t) t}$. However, the rate of growth, quantified by $\lambda_D(t)$ is in general different from $\lambda_L(t)$, due to the difference in the order of averaging. A straightforward application of Jensen's inequality \cite{jensen1906} gives $\lambda_L (t) \leq \lambda_D(t)$ at any finite time where the two values become equal in the limit $t\to\infty$ as the width of the distribution of $\lambda({\bf S},t)$ decreases as $t^{-2/3}$ (see below).

Fig.~\eqref{DxtVst} compares the numerical results of the linearised and non-linear equations of motion, which confirms the above expectations. In the limit $t\to\infty$, we find from linear extrapolation of our data $\lambda_L(\infty)=\lambda_D(\infty):=\lambda\approx 0.492(5)$.  This compares well with the value of $\lambda\approx 0.47$  reported earlier  \cite{de2012largest}. For any small but finite $\varepsilon$, $D(x,t)$ would eventually saturate to the value $1$, when de-correlation is complete (see Eq.~\eqref{Dreln}). The time for saturation goes as $\sim -\ln{\varepsilon}/\lambda$. Hence, the exponential growth-regime  lasts longer for smaller $\varepsilon$.  This can  be seen in Fig.~\eqref{DxtVst}  where we also plot the results from the non-linear dynamics for values of $\varepsilon=10^{-4}, 10^{-6}$ and $10^{-8}$. The inset shows that for the linearized dynamics, the scaling form in Eq.~\eqref{Dform} holds accurately over the entire time range, with $\mu \approx \lambda_D$. This means that we can identify a velocity dependent Lyapunov exponent through the relation $D(x=vt,t)\sim e^{2 \mu(v) t}$ with $\mu(v)=\lambda_D [1-(v/v_b)^2]$ to a very good approximation.  For the non-linear dynamics, as seen in Fig.~\eqref{DxtVsx}, the scaling form holds only for $x \sim v_b t$. 

\begin{figure}
 \centering
\includegraphics[scale=0.22]{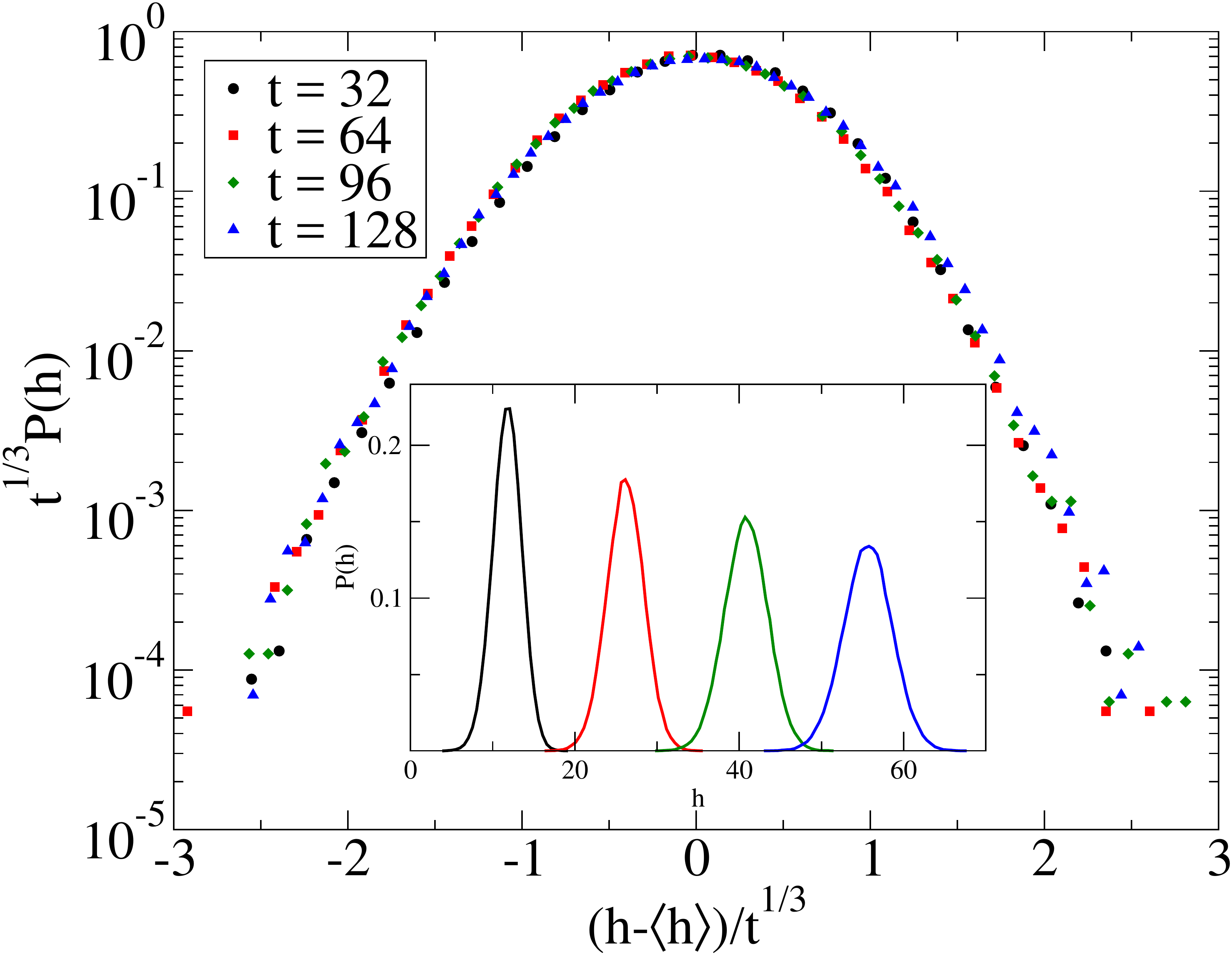}
\caption{(color onine) Distribution of the ``height'' variable $h(x,t)= \log [\delta {\bf S}^2(x,t)/\epsilon^2]/2$ at $x=0$. The inset shows the distribution of $h(0,t)$ at different times while the main plot shows the collpase of data obtained after a $t^{1/3}$ scaling.} 
\label{htdist}
\end{figure}

We now turn to the issue of realisation to realisation fluctuation of the wave-front and the finite variance in the arrival times, $t_{D_0}$, at a given $x$ (Fig. \eqref{tDvsx}). We define 
\begin{align}h(x,t)= \lim_{\epsilon \to 0} \log [\delta {\bf S}^2(x,t)/2 \epsilon^2]/2
\end{align}
(where we no longer average over initial configurations) and calculate $h(x,t)$ using the linearised equation of motion (Eq. \eqref{lindyn}) for ${\bf z}_x$. 
Our results so far suggest that 
\begin{equation}
h(x=vt,t)  =  t \mu(v)+ t^{1/3} \eta(x,t)~,
\end{equation}
where $\mu(v)$ is the velocity-dependent Lyapunov exponent, 
and $\eta$ describes the fluctuations.     In  Fig. \eqref{htdist}  we see that the probability distribution of $h(0,t)$ shows a clear $t^{1/3}$ scaling as mentioned above. 

The above observation leads us to interpret the dynamics of $h(x,t)$ as similar to the problem of interface growth \cite{pikovsky1994roughening} with $h(x,t)$ as the ``height function".  In particular, our numerical results for both $h(x,t)$ and $D(x,t)$ are consistent with the growth of height, as predicted from the KPZ equation for the so-called ``wedge'' initial conditions \cite{prolhac2011height}.  This would then suggest that the variable $\eta$  
follow a Tracy-Widom distribution. However, our system should differ from KPZ in that the noise from the chaos should have power-law correlations in time due to the diffusing conserved energy and magnetization densities.  The distributions in Fig. \eqref{tDvsx} inset and Fig. \eqref{htdist} are found to be more symmetric than Tracy-Widom and closer to Gaussians.  The reasons for this are at present unclear.

\paragraph*{Summary :} We have studied the butterfly effect in a classical Heisenberg spin chain at infinite  temperature and have shown that a systematic understanding of this effect includes two simultaneous, but logically complementary aspects -- the exponential growth and ballistic spread of an  infinitesimal local perturbation determined by the Lyapunov exponents and the butterfly speed. Both  effects are quantified by an appropriately defined measure that is naturally related to the OTOC recently studied in context of scrambling in quantum many-body systems \cite{maldacena2016bound, PhysRevLett.70.3339, kitaev,rozenbaum2017lyapunov,tsuji2017bound, swingle2016measuring, lucas2017energy, roberts2016lieb, iyoda2017scrambling, kurchan2016quantum,patel2017quantum,zhang2018information}. Though we have presented infinite temperature results, the above features of the butterfly effect  survive at finite $T/J\gg 1$~. We have obtained the scaling-form of the fluctuations of the propagation front via the KPZ model for interface growth. 
Notably, the above ballistic spread of perturbation is present even while the usual two-point dynamic spin correlator shows diffusion and hence does not reflect correlations spreading with the butterfly speed. A natural question then pertains to the nature of correlators that are directly sensitive to this ballistic effect. A closely related desideratum is an analytical derivation of the equation of motion for the propagating ballistic front. The features reported here for the nearest neighbour spin-chain are expected to survive in presence of further neighbour couplings, albeit, with different values for $\lambda$ and $v_b$. Such issues and particularly the effect of long-range spin exchanges form interesting future avenues of research, particularly the latter where the ballistic effects may not survive.

\begin{acknowledgments}
The authors acknowledge A. Baecker, S. Banerjee, R. Basu, J. Bec, C. Dasgupta, D. Dhar, R. Govindarajan, M. Kastner, V. Khemani, J. Kurchan, S. Lepri, S. N. Majumdar, A. Nahum, A. Politi, A. Polkovnikov, S. Ramaswamy and S. Sabhapandit for useful discussions. S. B. and R. M. acknowl- edge MPG for funding through the partner group of strongly correlated systems at ICTS. SB acknowledges MPIPKS for hospitality.  A. Dhar and AK acknowledge support of the Indo-French Centre for the promotion of advanced research (IFCPAR) under Project No. 5604-2. AK, SSR and SB acknowledges the support of SERB-DST (India) for project grants ECR/2017/000634,  ECR/2015/00036 and ECR/2017/000504 respectively. A. Dhar acknowledges support from grant  EDNHS ANR-14-CE25-0011 of the French National Research Agency. The numerical calculations were done on the cluster {\it Mowgli} and {\it zero} at the ICTS-TIFR. 
\end{acknowledgments}

\bibliography{biblio}
\end{document}